\documentclass{moriond}
\usepackage{amsmath}

\def\be{\begin{equation}}
\def\ee{\end{equation}}
\def\bea{\begin{eqnarray}}
\def\eea{\end{eqnarray}}

\newcommand{\Photo}{\includegraphics[height=35mm]{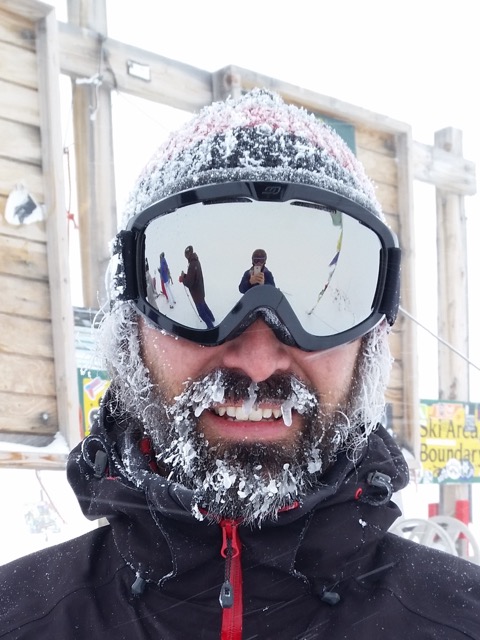}}

\begin{document}
\vspace*{4cm}
\title{QCD and High Energy Interactions: Moriond 2016 Theory Summary}

\author{Thomas Becher}

\address{Albert Einstein Center for Fundamental Physics, Institut f\"ur Theoretische Physik, Universit\"at Bern,
  Sidlerstrasse 5, CH-3012 Bern, Switzerland}

\maketitle\abstracts{
I summarize the theory talks presented at the 50 year anniversary conference.  The talks covered a wide range of topics, but common threads included next-to-next-to-leading order predictions for many $2\to 2$ processes, beyond QCD lattice simulations, work on the set of persistent flavor physics anomalies, and discussions of the significance and possible explanations of the diphoton excess at $750\,{\rm GeV}$.}

\section{Introduction}

This year's conference marked the 50 year anniversary of the Moriond meetings. Having these conferences become a place where important new results are announced and at the same time preserving the informal Moriond spirit is a big achievement. Congratulations to the organizers! Thanks to some celebratory talks, I finally learned where Moriond is and I enjoyed hearing about the early conferences and how the physics landscape looked at the time. Since many of the changes in our field are adiabatic, they are not always noticeable as they are happening and a longer-term perspective can be quite useful to realize how much has changed.

Given the short duration of the talks at the conference, speakers can only summarize their work. When reporting on  the 38 theory talks given at this year's conference, I will therefore necessarily violate the rule that one should not summarize summaries.  The wide array of topics covered makes it challenging to say something coherent, but on the plus side the large number of talks provides some statistics to identify new developments in the field and common threads connecting the work presented by different participants. 

\section{BSM Excesses}

The excess in the diphoton rate at $750\,{\rm GeV}$ presented by ATLAS \cite{ATLAS-CONF-2015-081} and CMS \cite{CMS:2015dxe} last December was dominating not only the session on Beyond the Standard Model (BSM) physics, but also the dinner-table discussions at the conference, and the new experimental results \cite{ATLAS-CONF-2016-018,CMS:2016owr} presented at the Moriond conferences added to the excitement. While no additional data was collected, the further analyses performed since December all somewhat strengthened the excess. For ATLAS, the new results included a dedicated analysis for the spin-2 hypothesis with looser cuts, as well as a reanalysis of the $8\, {\rm TeV}$ data and a check on their compatibility with the excess observed at $13\, {\rm TeV}$. CMS managed to add $0.6\,{\rm fb}^{-1}$ of data collected when their magnet was switched off. They increased their sensitivity by 10\% and their mass resolution by 30\%. The largest excess is now also at $750\,{\rm GeV}$, while the December results had  preferred an invariant mass of $760\,{\rm GeV}$.

There were strong, diverging opinions voiced at the conference as to whether the excess is a statistical fluctuation or a new particle. As a general rule, given the high stakes, the experimental speakers and participants emphasized the low global significance of the individual results, which are $2.0\sigma$ for ATLAS and $1.6\sigma$ for CMS (with $8$ and $13\,{\rm TeV}$ results combined), while theorists tended to stress the fact that both experiments see the excess and focus on the highest local significances reported, i.e.\ $3.9\sigma$ at  $13\,{\rm TeV}$ and $2\sigma$ at $8\,{\rm TeV}$ from ATLAS, and $3.4\sigma$  for CMS (with $8$ and $13\,{\rm TeV}$ results combined). The good thing is that we won't have to wait much longer to know more. After the conference ended, the LHC has successfully resumed running and by now it has already collected more than three times the data of last year's run. If the resonance is really there, we should have confirmation by the time of summer conferences. 

{\em If} this excess turns into a discovery, it is difficult to overstate its importance and I agree with Abdelhak Djouadi's statement that it would be the ``most important discovery since decades''. It would be the observation of the first particle beyond the Standard Model (SM) and would likely be followed by a plethora of additional states in the not too distant future, as stressed by Alessandro Strumia in his overview talk.\cite{Strumia:2016wys} This would finally provide guidance on how to extend the SM and it would be extremely interesting to understand the role of the newly discovered states in addressing some of the shortcomings of the SM (dark matter, strong CP problem, naturalness, \dots). Given these prospects, it is not astonishing that the excess caused tremendous excitement and prompted a lot of work on possible explanations. As a result, more than 350 theory papers with possible explanations were written, demonstrating that such a resonance can be accommodated in many New Physics scenarios even though such a signal was generally not anticipated. Alessandro Strumia gave a very nice overview over the different options  and I refer to his contribution to the proceedings for more details and references.\cite{Strumia:2016wys} Concerning the different possible explanations, it's good to keep in mind that at this point we know precious little about the new particle (assuming it is there).  We only know the mass and production cross section times branching fraction to $\gamma\gamma$ (the width could be large or small). Gauge invariance then implies that it also couples to $Z$-bosons, so it will be important to search for a signal in $Z\gamma$ and $ZZ$ production (with a suitable coupling to $W$-pairs, one could suppress one of these channels but not both). The most minimal option was discussed by Tanumoy Mandal, who checked whether it is possible to explain the resonance in a model, where the new particle does not couple to gluons and quarks but is produced by colliding photons from the initial state.\cite{Danielsson:2016nyy} This leads to tensions with the $8\,{\rm TeV}$ data since the number of photons at $13\,{\rm TeV}$ is not much larger than at $8\,{\rm TeV}$. Indeed, parton luminosity considerations make a production from gluon or heavy-quark fusion most likely. 

Let me briefly go over the other talks in the BSM session. Carlos Wagner discussed three versions of the NMSSM, aptly named the good, the bad and the ugly. The good news is that the NMSSM can provide the necessary alignment to explain the good agreement of Higgs physics with the SM.\cite{Carena:2015moc} With some tuning (the bad), it can also provide the enhancement of the Higgs rate in $t\bar{t} h$ observed in Run I.\cite{Badziak:2016exn} Finally, allowing for $R$-parity violation (the ugly) also the flavor anomalies, together with the $750\,{\rm GeV}$ excess can be accommodated.  Brian Shuve discussed collider signatures from hidden sectors and long-lived particles. There are many different models in this category and they yield interesting signals such as displaced jets. Finally, Alexander Belyaev discussed dark-matter searches at the LHC in NSUSY \cite{Barducci:2015ffa} and within the EFT framework.

\section{Flavor Puzzles}

\begin{figure}
\begin{minipage}{0.49\linewidth}
\centerline{\includegraphics[height=0.8\linewidth]{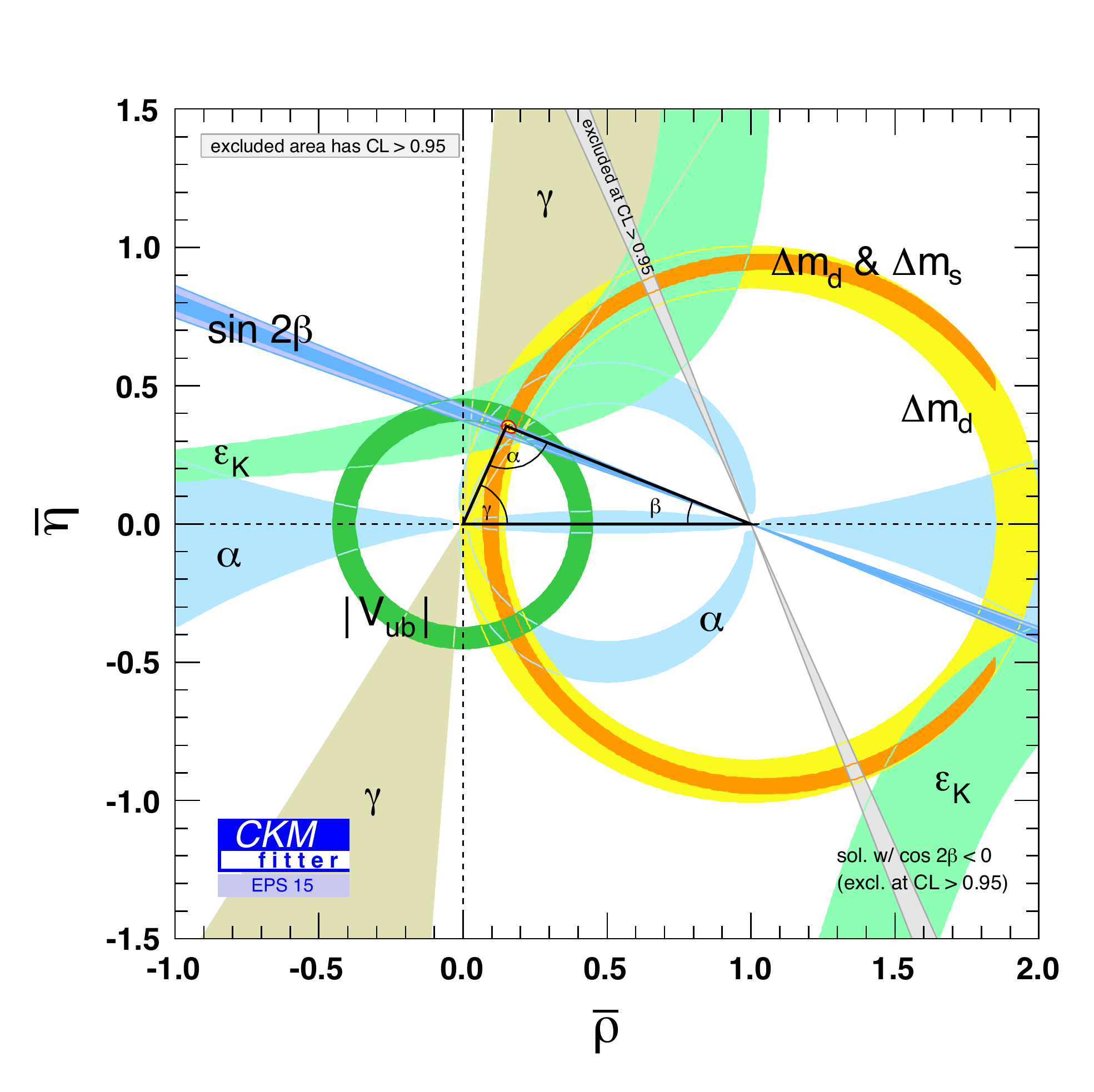}}
\end{minipage}\hfill
\begin{minipage}{0.43\linewidth}
\centerline{\includegraphics[height=0.8\linewidth]{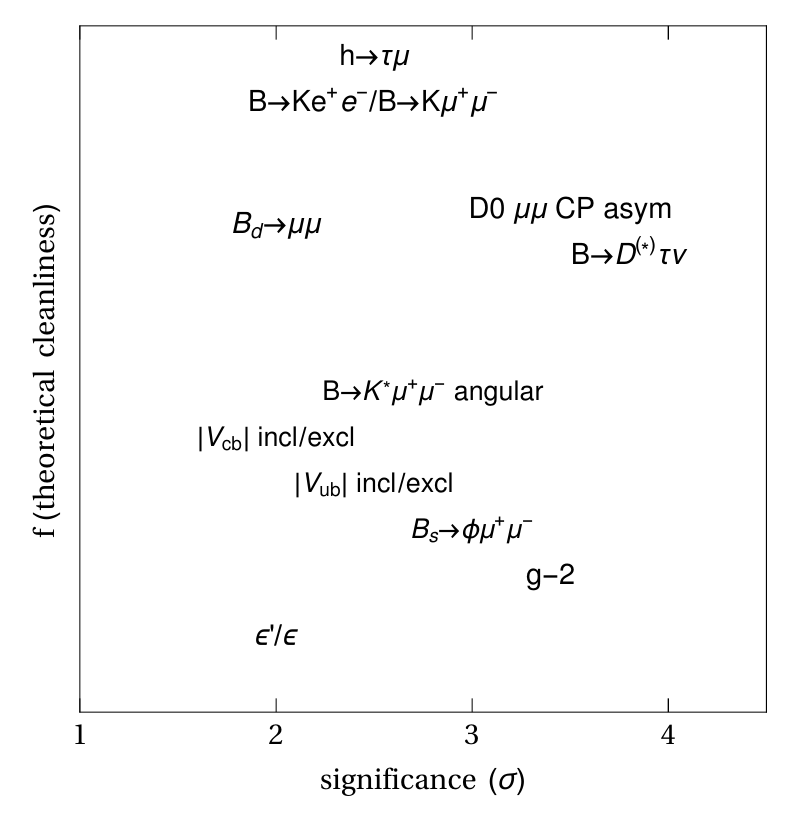}}
\end{minipage}
\caption[]{Left: Status of the CKM fit. Right, from Z.\ Ligeti: the different flavor physics anomalies as ``an undefined function of an ill-defined variable: the theoretical cleanliness.''}
\label{fig:tensions}
\end{figure}

\begin{figure}
\begin{minipage}{0.44\linewidth}
\centerline{\includegraphics[height=0.8\linewidth]{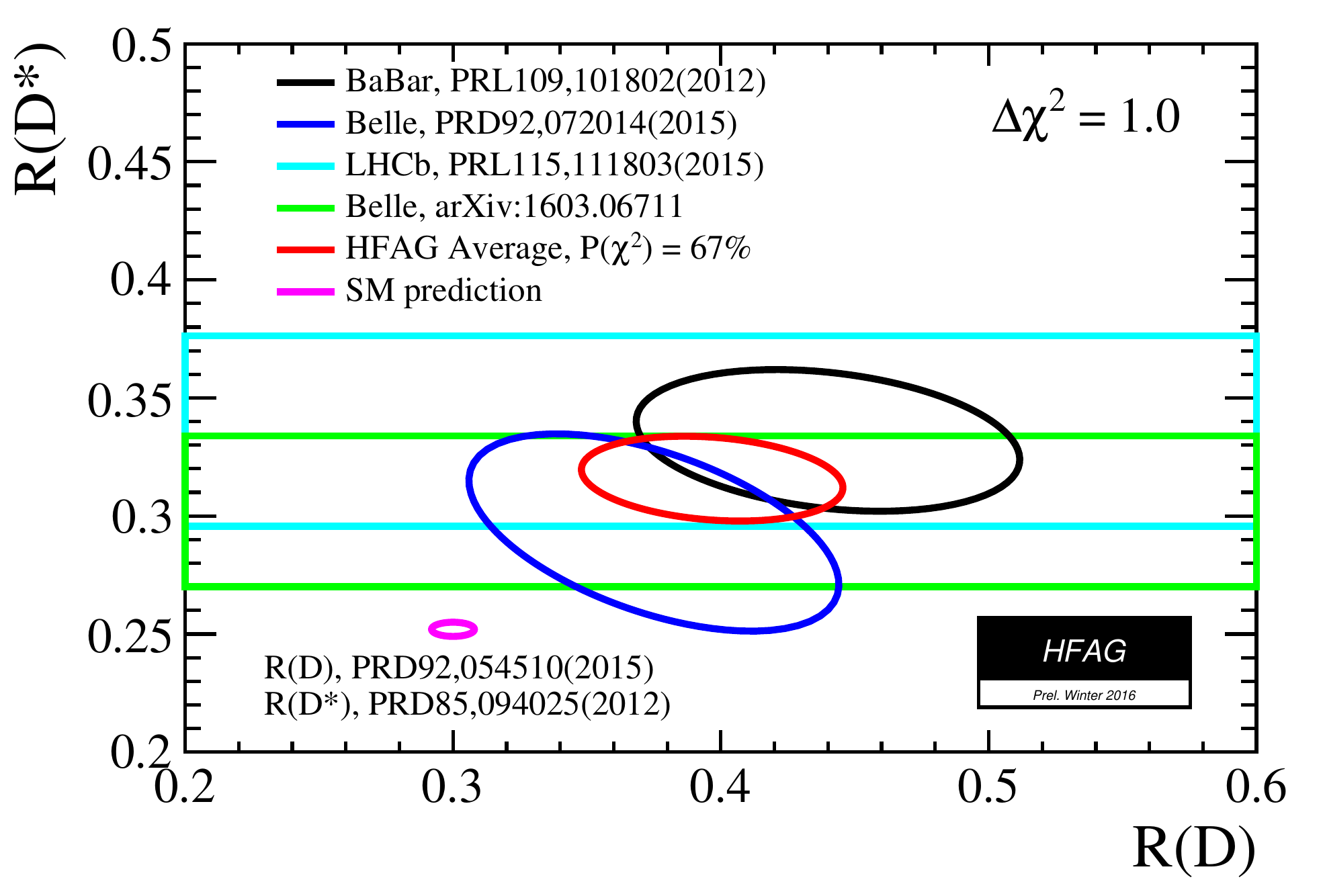}}
\end{minipage}\hfill
\begin{minipage}{0.43\linewidth}
\centerline{\includegraphics[height=0.8\linewidth]{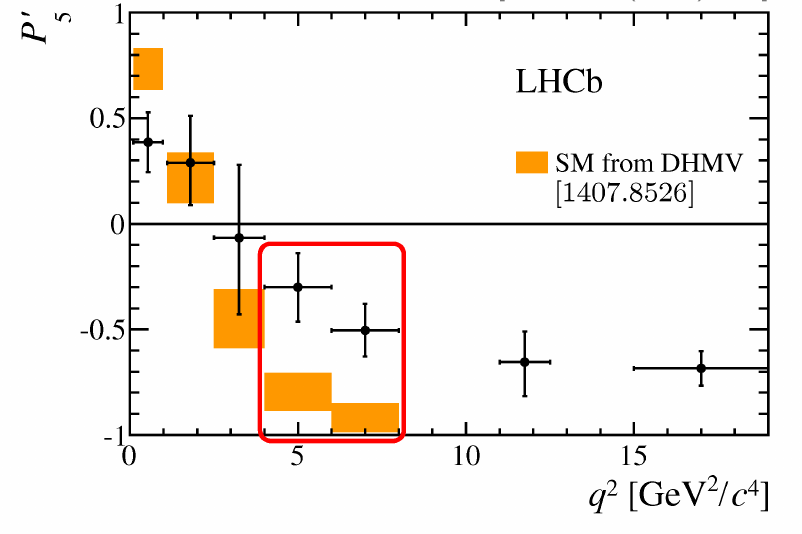}}
\end{minipage}
\caption[]{Left: Ratios $R(D)$ and $R(D^*)$. Right: Form factor ratio $P_5'(q^2)$.}
\label{fig:RDP5}
\end{figure}

The flavor session started with an overview talk by Zoltan Ligeti. He reminded us that the commissioning of SuperKEKB is ongoing and that the Belle II experiment will start soon. The design luminosity of this new machine is 40 times larger than the luminosity achieved by the KEKB collider. Also, NA62 is scheduled to have around 200 run days this year which would translate into about 50 events in $K^+ \to \pi^+ \nu \bar{\nu} $ at SM level. These new, larger data sets are certainly welcome since flavor physics is in an interesting quantum-mechanical superposition. On the one hand, the CKM picture of CP violation has been confirmed with good precision, see the left plot in Figure \ref{fig:tensions}, but at the same time, there are currently several interesting, and in some cases persistent, anomalies observed, which were the focus of the talks at the conference. They include deviations from the SM predictions in 
\begin{itemize}
\item $R(D^{{\scriptscriptstyle (}\scriptstyle *{\scriptscriptstyle)}})=\Gamma(B \to D^{{\scriptscriptstyle (}\scriptstyle *{\scriptscriptstyle)}} \tau \bar{\nu})/\Gamma(B \to D^{{\scriptscriptstyle (}\scriptstyle *{\scriptscriptstyle)}} \ell \bar{\nu})$ with $\ell = e,\mu$,
\item  $R(K)=\Gamma(B \to K \mu^+\mu^-)/\Gamma(B \to K e^+e^-)$,
\item angular distributions in $B \to K^* \mu^+\mu^-$,
\item $\Gamma(B_s \to \phi \mu^+ \mu^-)$
\item $g-2$ of the muon,
\item $h\to \tau \mu$.
\end{itemize}
The significance of each deviation is shown in Figure \ref{fig:tensions}, taken from Z.~Ligeti's talk. In the same plot he orders the deviations by ``theoretical cleanliness''. Flavor observables are mostly measured in low-energy processes, and strong-interaction effects can make it difficult to arrive at theoretical predictions with controlled, small uncertainties. However, there are a variety of theoretical tools which can make precise computations possible. For example, the $B \to D $ transition form factor can be accurately computed using lattice methods \cite{Lattice:2015rga,Na:2015kha} and heavy-quark symmetry relates the $B \to D $ and $B \to D^*$ transitions so that the ratios $R(D)$ and $R(D^*)$ are predicted with high precision. The nature and size of the error bars makes it unlikely that QCD effects can explain the observed $4 \sigma$ deviation shown in the left panel in Figure \ref{fig:RDP5}. (Recently, a method was proposed \cite{Bordone:2016gaq} to also check QED corrections which the experiments include using PHOTOS; at least for  $R(K)$, they seem to be well modeled.\cite{Gratrex:2015hna}) It would be somewhat ironic if New Physics would be observed in tree-level transitions since the main selling point of flavor physics is that it has great sensitivity to New Physics in rare processes, which are forbidden at tree level in the SM.  

Flavor Changing Neutral Current (FCNC) transitions are a prime example of such rare processes. Interestingly, in $B \to K^* \mu^+\mu^-$ a  deviation is observed in the ratio $P_5'$  of form factors displayed in the right plot in Figure \ref{fig:RDP5} showing results from LHCb.\cite{Aaij:2015oid} The two bins marked in red show a deviation of $2.8\sigma$  and $3.0\sigma$. The theoretical description of the $B \to K^*$ transition is far more involved than for the ratios $R(D^{(*)})$. Lattice computations cannot be used in the low-$q^2$ region where the deviation is observed. Instead one relies on form factor relations in the heavy-quark limit and tries to reduce QCD corrections by using form factor ratios such as $P_5'$.\cite{DescotesGenon:2012zf} The proper way to analyze New Physics effects is to use a low-energy effective Lagrangian and the most interesting operators in this context are 
\begin{align}
O_7 & \propto m_b \,\bar{s} \sigma_{\mu\nu} P_L b \,F^{\mu\nu}\,,  & O_9 & \propto \bar{s} \gamma_\mu P_L b \, \bar{\ell} \gamma^\mu \ell \,,& O_{10} & \propto \bar{s} \gamma_\mu P_L b  \, \bar{\ell} \gamma^\mu\gamma_5 \ell\,.
\end{align}
In the SM only left-handed operators are present, $P_L=\frac{1}{2}(1-\gamma_5)$, but New Physics can also induce opposite chirality operators. Several groups have performed global fits of $b\to s \ell^+\ell^-$ transitions, the ones published during the last year include \cite{Beaujean:2015gba,Descotes-Genon:2015uva,Ciuchini:2015qxb,Meinel:2016grj}. Christoph Bobeth presented a thorough review of the fits and the theoretical assumptions that go into them. He stressed that these fits are ``serious business'' and that it is necessary to look at the fine print when assessing and comparing results. The main issue one faces is that the fit must distinguish  corrections to the heavy-quark limit from New Physics contributions. Unfortunately, these power corrections can only be estimated. The fits leave the size of the corrections open, but rely on the form obtained using model calculations. One then fits New Physics contributions to the Wilson coefficients of the operators together with hadronic and other input parameters. A common feature of the fits is that the best fit point is obtained with a new- physics contribution $C_9^{\rm NP} \approx -0.25 \, C_9^{\rm SM}$ while the other operators can be kept at their standard-model values, as first observed in \cite{Descotes-Genon:2013wba}.  The deviation in $R(K)$ can be accommodated by allowing for lepton-flavor dependent Wilson coefficients. In the fit of Descotes et al.\ \cite{Descotes-Genon:2015uva}, the SM hypothesis has a pull of $4 \sigma$ for $\ell = \mu$ and even $5 \sigma$ for $\ell = e,\mu$ (the $p$-value of the SM hypothesis is 16\%). 

It is interesting to ask what kind of New Physics could give rise to the pattern of anomalies we currently observe. Andreas Crivellin, discussed different models which can explain some or all of the above deviations. The deviations in $b\to s \ell^+\ell^-$ can be explained, for example, by a $Z'$. If one wants to explain $R(D^{{\scriptscriptstyle (}\scriptstyle *{\scriptscriptstyle)}})$, charged Higgs bosons or leptoquarks are an option. Indeed, a suitable leptoquark model can simultaneously explain the deviations in both observables as well as the deviation in the muon $g-2$. For a detailed discussion and a list of references proposing different explanations, I refer the reader to Andreas Crivellin's contribution to the proceedings.\cite{Crivellin:2016ivz} One comment I cannot resist to make is that one should carefully choose which anomalies one wants to  explain and which ones are better left alone because they will eventually go away. The preferred non-zero value for ${\rm Br}(h\to \tau \mu)$ from CMS \cite{Khachatryan:2015kon} and ATLAS  \cite{Aad:2015gha}, for example, is not  significant (combined the deviation is a $2.6\sigma$ effect) and looks too large to be true. Indeed, new preliminary CMS results \cite{CMSSLIDES} seem to lower the significance even further.

The only theory talk in the flavor session not directly concerned with deviations from the SM was given by Cai-Dian L\"u who presented a global analysis of $B\to D M$ decays, where $M$ is a light meson.\cite{Zhou:2015jba} His analysis relies on factorization and $SU(3)$ relations (using the  topological amplitude approach). After fixing the input parameters using a subset of decay modes, the branching ratios obtained for the more than one hundred decays in this category seems to match quite well to the available experimental results and provide predictions for the ones which have yet to be measured.

\section{Perturbative QCD: the year of NNLO}

One thing I found striking at this year's conference was the large number of results for $2 \to 2 $ collider processes at next-to-next-to-leading (NNLO) order in perturbation theory. Indeed, at this point the list of NNLO predictions for $2\to 2$ processes is almost complete (an important missing item is the 2-jet cross section, for which partial results are available \cite{Ridder:2013mf}) and public codes to compute fiducial cross sections for a variety of these processes are becoming available. 

In the past few years, the majority of talks were concerned with NLO computations and their automatization. There has been tremendous progress in this area and we have reached a stage where matched NLO computations for simple processes are easily available through public codes. Theory efforts in this area now focus on the automation of NLO EW effects, which are generically of similar size as NNLO QCD effects but are enhanced by Sudakov logarithms at high energies, and on the inclusion of decays of unstable particles and off-shell effects, which are small in inclusive cross sections, because they are suppressed by $\Gamma/m$, but can be large in distributions. The relevance of off-shell effects was discussed by Malgorzata Worek who presented the NLO result for top pair production in association with a jet, including the top and $W$ decays.\cite{Bevilacqua:2015qha,Worek:2016qqb} Including the decays, one ends up with the process $pp \to e^+\nu_e \mu^- \bar{\nu}_\mu b \bar{b} j$. The resulting computation involves heptagon diagrams and is obviously very challenging. Together with Mike Trott's result on $h\to \gamma\gamma$ at one-loop including $d=6$ operators in the SM effective field theory \cite{Hartmann:2015aia}, these were the only talks at NLO, all other speakers were presenting NNLO results. 

\begin{figure}
\begin{minipage}{0.44\linewidth}
\centerline{\includegraphics[height=0.8\linewidth]{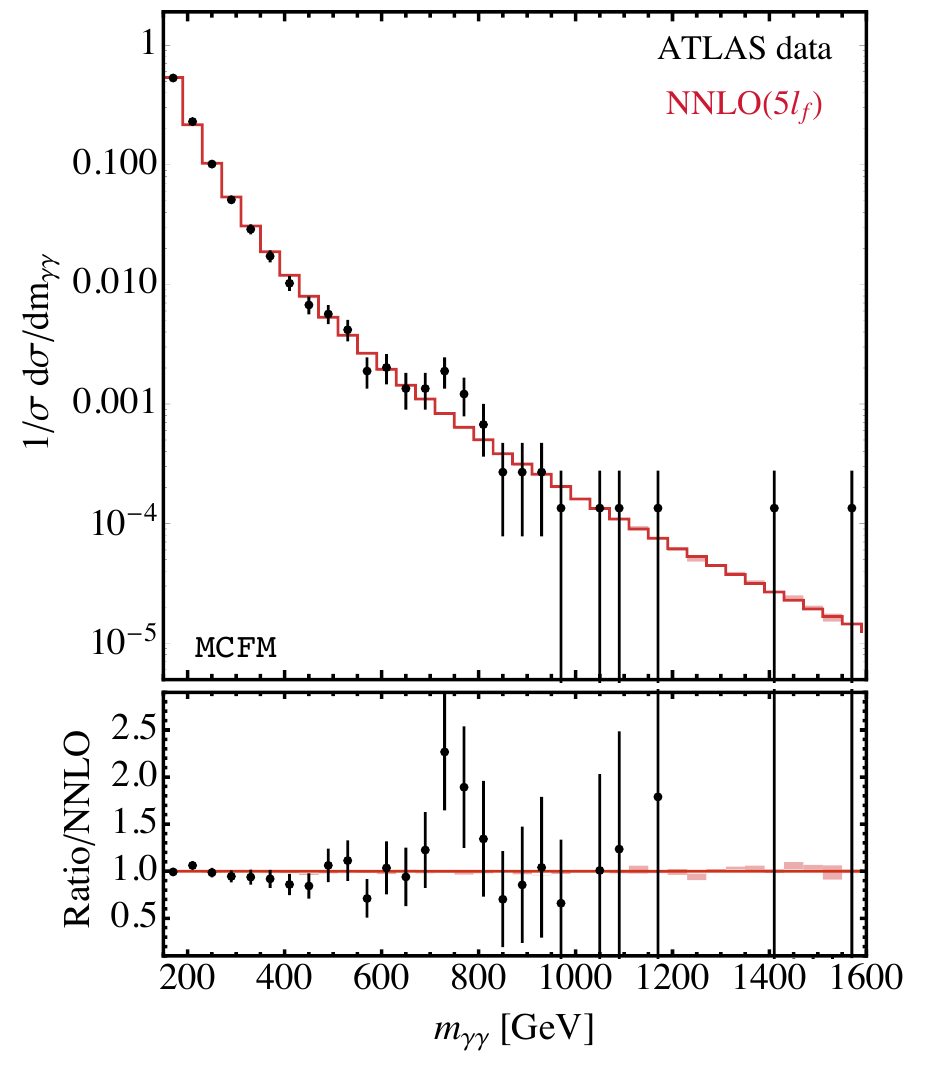}}
\end{minipage}\hfill
\begin{minipage}{0.43\linewidth}
\centerline{\includegraphics[height=0.8\linewidth]{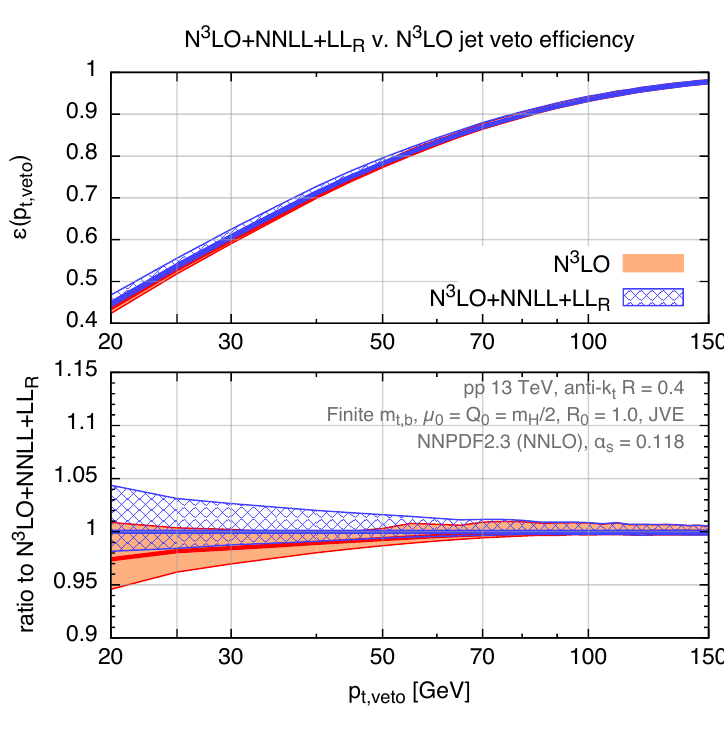}}
\end{minipage}
\caption[]{Left: Diphoton production at NNLO \cite{Campbell:2016yrh} compared to ATLAS data. Right: Higgs cross section with a jet veto.\cite{Banfi:2015pju}}
\label{fig:NNLO}
\end{figure}

The progress in NLO computations is often celebrated as the ``NLO revolution'' since it is based in part on methods which are radically different than the traditional way of doing such computations and which tame the strong increase in complexity for higher-multiplicity processes. The avalanche of NNLO results, on the other hand, seems to be based on clever ideas improving and extending existing technology. Key ingredients for the new results are improved methods for solving differential equations for two-loop integrals \cite{Henn:2013pwa,Henn:2014qga} and new mathematical techniques to handle the resulting iterated integrals.\cite{Duhr:2014woa} Using these techniques the two-loop amplitudes were obtained for arbitrary massive vector bosons in the state \cite{Henn:2014lfa,Caola:2014iua,Caola:2014lpa,Papadopoulos:2014hla,Gehrmann:2015ora,vonManteuffel:2015msa} and implemented in a public code.\cite{VVamp} Not only the two-loop integrals, but also the real-emission corrections are very challenging at NNLO and also in this area there has been a lot of recent progress. Interestingly, the results presented at the conference were computed using a variety of   methods for their computation, namely
\begin{enumerate} 
\item Antenna subtraction \cite{GehrmannDeRidder:2005cm} (for $pp \to Z+j$, talk by Alexander Huss),
\item\label{stripper} Sector improved residue subtraction \cite{Czakon:2010td} (for $pp \to t\bar{t}$, talk by David Heymes),
\item \label{slicing} Slicing: $q_T$-subtraction \cite{Catani:2007vq} (for $pp\to \gamma\gamma$, $pp\to W\gamma$, $pp\to ZZ$, $pp\to WW$, talk by Stefan Kallweit) and $N$-jettiness subtraction \cite{Boughezal:2015dva,Gaunt:2015pea} (for $pp\to \gamma\gamma$, $pp\to W H $, $pp\to Z H $, $pp\to VH$ followed by $H\to WW$, talk by Ciaran Williams),
\item Projection to Born \cite{Cacciari:2015jma} ($pp \to h$ in vector-boson fusion, talk by Alexander Karlberg).
\end{enumerate}
(A fifth method, CoLoRFulNNLO, recently produced first results in $e^+e^-$.\cite{DelDuca:2016csb,DelDuca:2016ily}) It is interesting that all these quite different approaches now produce phenomenological results, despite the fact that some of them have been introduced only during the past year. The above list only shows results presented at the conference, it is therefore incomplete and suffers from  some selection bias (for example, it is missing the $pp \to H+j$ \cite{Boughezal:2013uia,Boughezal:2015dra} results obtained using a version \cite{Boughezal:2011jf} of Method \ref{stripper}). Nevertheless is interesting to observe that the largest number of results was obtained using slicing methods. These methods use a variable $v$ (concretely $q_T$ or $N$-jettiness) which vanishes in the purely virtual limit. If the variable is non-zero, one therefore has at least one emission and  can use existing NLO codes to compute the observable. In the small-$v$ region, on the other hand, one can simplify the problem by expanding in $v$ and using the inclusive cross section. The fact that one can recycle NLO results and codes is, I think, an important reason for the high rate with which $2\to2$ NNLO results were produced using Methods \ref{slicing} during the last year. The same fact greatly helps automatizing these computations. Stefan Kallweit discussed in his talk this automatization for $V V'$ production using $q_T$-subtraction. The NLO computations for $q_T \neq 0$ are based on the OpenLoops framework.\cite{Cascioli:2011va} The variable $q_T$ can only be used for slicing if the final-state particles are color neutral. To also deal with with jets in the final state, one can use an event-shape variable as we pointed out in \cite{Becher:2015gsa}. $N$-jettiness \cite{Stewart:2010tn} is an extension of thrust to processes with several jets and can serve for this purpose.  By computing $pp \to W+j$ it was demonstrated \cite{Boughezal:2015dva} that this method is indeed practical. Ciaran Williams presented several computations based on this method and discussed progress in producing a version of MCFM which includes NNLO for selected processes. A new version of this code with NNLO for color-singlet processes was indeed released shortly after the conference \cite{Boughezal:2016wmq}, among them diphoton production which is shown in Figure \ref{fig:NNLO}. The plot shows that the NNLO predictions agree nicely with the ATLAS measurement, except for an excess in the region around $m_{\gamma\gamma} \approx 750\,{\rm GeV}$. Ciaran Williams was stressing the importance of computing rather than fitting backgrounds. Backgrounds are not necessarily smooth and there is not always enough side-band data to reliably fit the them. However, while the $m_{\gamma\gamma}$ spectrum indeed has some features, in particular near $m_{\gamma\gamma}=2\, m_t$, these are per-cent level effects and a fit with a smooth function works well.

Given the large number of NNLO results presented at the conference, I will not comment on them individually, but a common feature is that in all cases the corrections lead to a better agreement with data. Also, as expected, their inclusion reduces the scale variation. At the same time, it is not uncommon that the NNLO corrections lie outside the NLO scale variation bands, highlighting the well-known fact that these bands are a relatively crude way of estimating theoretical uncertainties. Interestingly, for the few cases where they are known, namely Higgs production in gluon fusion (with \cite{Banfi:2015pju} or without \cite{Anastasiou:2015ema} a jet veto) and in vector boson fusion \cite{Dreyer:2016gwc} the N$^3$LO order corrections do lie inside the NNLO bands. Of course, whether scale-variation bands overlap depends on the default value of the scale which one adopts and it is not always clear what value one should adopt in a given process. For top production David Heymes discussed this issue in detail in his talk. He presented differential predictions for top production \cite{Czakon:2015owf,Czakon:2016ckf,Czakon:2016dgf} and demonstrated that dynamical scales can significantly improve the apparent convergence for quantities like the invariant mass of the top-pair or the transverse momentum of the top. 

A crucial ingredient for the computation of any hadron-collider process are parton distributions (PDFs). In the mini-session on the topic Stefano Carrazza presented the PDF4LHC15 recommendation, a new PDF combination of the PDF4LHC working group.\cite{Butterworth:2015oua} The new combination is based on sets which include Run I data and is more convenient and meaningful than the earlier envelope prescription (though not universally accepted \cite{Accardi:2016ndt}). The most important ingredient to PDF fits are DIS data and Volodymyr Myronenko presented the final HERA combination and associated HERAPDF2.0 set.\cite{Andreev:2014wwa} Aleksander Kusina presented nCTEQ15, a new nuclear PDF set.\cite{Kovarik:2015cma} Finally, Brian Page and Inseok Yoon presented results for polarized proton collisions obtained at STAR\cite{Adamczyk:2014xyw,Adamczyk:2014ozi} and PHENIX\cite{Adare:2015ozj}, which are for example interesting to study the spin-content of the proton.

\section{From fixed-order perturbation theory to resummation and back}

As discussed above, choosing the renormalization and factorization scales wisely is important, but if processes are sensitive to several disparate physical scales, there is no theoretically preferred scale choice and to be conservative one should simply vary the scales to cover the entire range of physical scales which are relevant in a given observable. To do better than this, one has to first factorize the physics at the different scales and then compute each part at its appropriate scale. By using evolution equations for the individual factors, one then manages to resum the terms which are enhanced by logarithms of the scale ratios. 

Important examples of multi-scale problems are jet (sub-)structure observables, which are playing an increasingly important role at the LHC. The main goal of such observables is to distinguish ordinary QCD jets from jets containing heavy quarks, electroweak bosons or BSM particles, which is difficult if the decay products are highly boosted. Zack Sullivan discussed $b$-tagging and presented a new $b$-tagger for boosted jets.\cite{Pedersen:2015knf,Sullivan:2016lpz} Jesse Thaler gave an overview over recent developments in the field of jet substructure. While most substructure studies are based on parton shower Monte-Carlo programs, there has been a lot of progress in understanding jet structure from first principle computations. He illustrated this point using the example of momentum balance among subjets. This is a collinear unsafe observable, but interestingly Sudakov suppression renders it calculable and the result has several interesting features.\cite{Larkoski:2015lea} Jesse and collaborators then used the CMS opendata \cite{opendata} to verify that these are indeed present in the experimental data. Frederic Dreyer showed how to resum the leading logarithms (LL) of the jet radius $R$ in jet processes.\cite{Dreyer:2016gwc,Dasgupta:2014yra,Dasgupta:2016bnd} Varying $R$ gives an interesting additional handle on systematic effects since hadronisation corrections are suppressed at large $R$, while underlying-event effects are suppressed at small $R$. The subleading log terms at NNLO turn out to be sizeable and he included those arising at NNLO by matching to fixed order. 

The talks at the conference made it clear that the relationship between fixed order and resummed computation can be quite synergistic. For example, the singular terms in the phase-space integrals arise from soft and collinear emissions which are precisely the regions responsible for the occurrence of large logarithms. The slicing methods discussed in the previous section are based on factorization theorems for the cross section when the slicing parameter goes to zero. Originally, these factorization theorems were derived to perform resummation, but now prove useful for NNLO fixed-order computations. Conversely, it is clear that the computation of the ingredients of the factorization theorems require fixed-order computations. The automated NLO fixed-order codes can therefore be used to perform automated resummation, as we did for jet-veto cross sections.\cite{Becher:2014aya} An important ingredient for resummations are the so-called hard functions and Wouter Waalewijn showed that by using a suitable operator basis, QCD on-shell amplitudes are equal to bare Wilson coefficients in SCET.\cite{Moult:2016lyu}

The most precise available predictions for collider processes combine higher-log resummation with fixed-order results. The result for the Higgs cross section in the presence of a jet veto presented by Pier Francesco Monni is a very impressive example of such a prediction. Using the N$^3$LO fixed-order result for the total Higgs cross section \cite{Anastasiou:2015ema} together with the NNLO result for Higgs in association with a jet \cite{Boughezal:2015dra}, one obtains the cross section in the presence of a jet veto to N$^3$LO. Pier Francesco and collaborators then combine this with NNLL resummations of logarithms of the jet veto scale and LL resummation of logarithms of the jet radius $R$ to obtain the highly accurate prediction for jet-veto efficiency shown in Figure \ref{fig:NNLO}.

Having discussed all these higher-order computations and higher-logarithmic resummations, I should stress that it is equally important to identify good observables, i.e. observables which have intrinsically small uncertainties and are suitable for precision studies. Marat Freytsis and collaborators have analyzed ratios of diboson cross sections.\cite{Frye:2015rba}  Analyzing the $SU(2)\times U(1)$ structure of these processes and its breaking, they are able to identify ratios, such as $\sigma(Z\gamma)/\sigma(\gamma\gamma)$, in which QCD uncertainties cancel to a large degree.

\section{Beyond QCD lattice gauge theory}

Another thing I found remarkable was that none of the talks in the lattice session was on lattice QCD. The only QCD talk of the session was by Anton Rebhan who was using the Witten-Sakai-Sugimoto model to compute glueball decays.\cite{Brunner:2015sba,Brunner:2015pxm,Rebhan:2016ecl} However, in this case the absence of evidence is not evidence of absence (to use one of the cliche sentences of which physics papers are full of). There is an increasing number of QCD physics results obtained using lattice simulations as witnessed by the fact that there is a working group dedicated to compiling and averaging these results.\cite{Aoki:2013ldr} However, the large increase in computer power and new powerful algorithms now make it possible to also study BSM theories, as stressed by Rich Brower in his talk. An interesting scenario to explore is whether the Higgs could be the dilaton of a near-conformal model. To address this question Anna Hasenfratz and collaborators have studied $SU(N)$ gauge theory with $N_f = N_l +N_h$ fermions. Above their mass scale, the additional $N_h$ heavy quarks slow the running of the coupling and can lead to a near-conformal model with a broken chiral symmetry for the $N_l$ light fermions.  The case $N_l +N_h = 4+8$ was explored using staggered fermions and this pilot study indeed observed a window where the coupling is running slowly, together with a relatively light $0^{++}$ state (which should then play the role of the Higgs).\cite{Brower:2014ita,Brower:2015owo} Of course, such a theory would involve a tower of resonances in the few-TeV range and observing such resonances experimentally  would be a strong motivation to explore such scenarios in detail. It is great to see numerical simulations of BSM theories, but it is fair to say that these simulations are still at the level of toy models trying to establish certain general properties of strongly coupled theories and that there is a long way to go to make contact with experiments. 

Dorota Grabowska presented a very interesting proposal for a lattice formulation of chiral gauge theories such as the SM.\cite{Grabowska:2015qpk} Thanks to 't Hooft and Veltman \cite{'tHooft:1972fi}, we have a perturbative expansion for chiral gauge theories, but it is of course important to also have a non-perturbative formulation since there can be effects which are not captured by perturbation theory. The basic problem is that  lattice formulations necessary include chiral doublers and one then needs a mechanism to decouple these mirror fermions in the continuum limit. For global chiral symmetries several formulations exist which achieve this, among them domain wall fermions.\cite{Kaplan:1992bt} The formulation of chiral gauge theory she presented is based on domain-wall fermions but uses gradient flow to extend the gauge field into the fifth dimension. The gradient flow dampens high-momentum gauge-field modes along the fifth dimension and effectively provides a soft form factor for these. To make the construction appear less intimidating for non-experts like myself Dorota calls the resulting mirror fermions ``fluffy". She stressed that at this point there are still open questions which need to be studied to prove that the formulation indeed has all the correct properties in the continuum limit. Once these are addressed it will be interesting to see what non-perturbative effects arise due to the presence of the fluffy mirror fermions.

\section{The  Riches and Ridges of Heavy-Ion Collisions}

\begin{figure}
\includegraphics[width=0.6\textwidth]{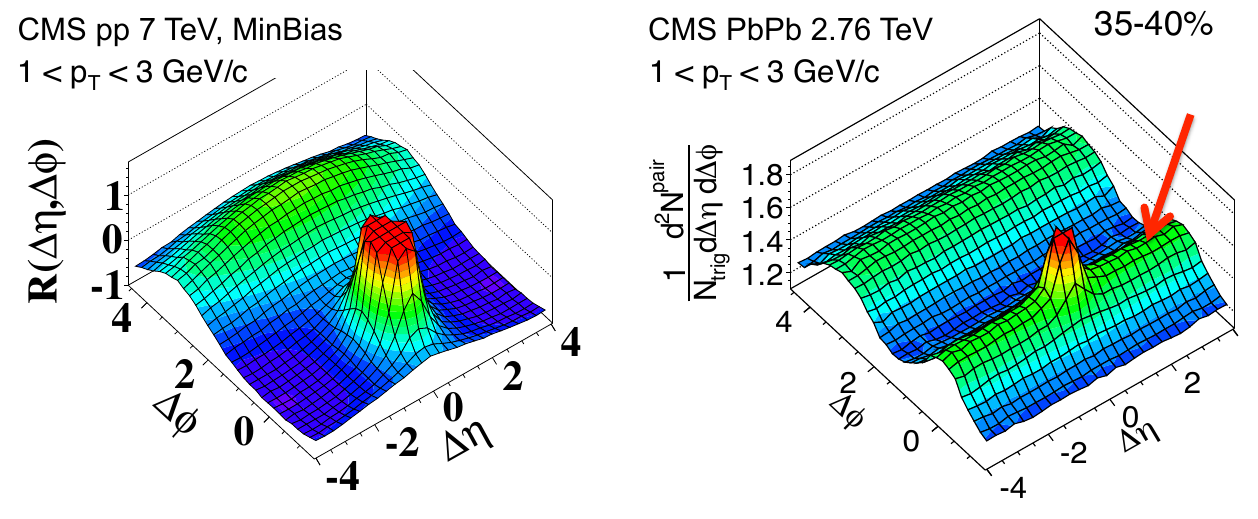}\hspace{1cm}\includegraphics[width=0.29\textwidth]{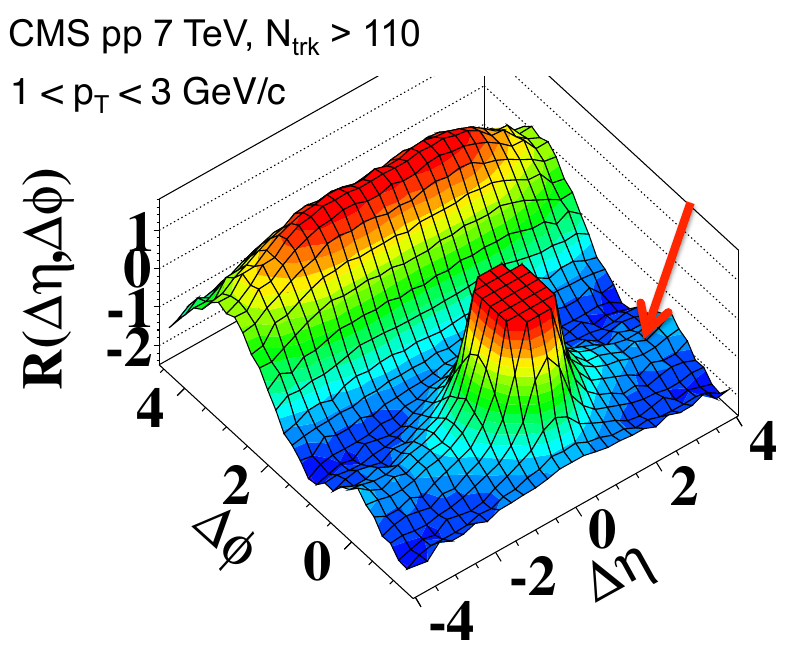}
\caption[]{Two-particle angular correlations in proton-proton and lead-lead collisions, from 
\cite{Khachatryan:2010gv}. The red arrow in the middle plot marks the ridge, two-particle correlations arising from the presence of the expanding medium. Interestingly, similar correlations are also present in high-multiplicity proton-proton events, see the right plot.}
\label{fig:ridges}
\end{figure}

The full complexity and richness of QCD is at play in heavy-ion collisions, which were the topic of the last day of the conference. Jan Fiete Grosse-Oetringhaus gave a great overview of the current status and discussed the different available probes of the Quark-Gluon Plasma (QGP). These include particle flow and particle correlations, which measure global properties of the expanding medium after the collision, as well as the modification of hard probes by the medium. Among the hard probes are jet quenching and quarkonium production. A common theme among the theory presentations was that the different physics (hard QCD, QGP, cold nuclear effects, hadronisation, \dots) in heavy-ion collisions can be difficult to disentangle, since it can produce the same phenomena.

A case study for this was presented by Elena Ferreiro who discussed quarkonium suppression and its interesting history. Around 2000, the suppression of quarkonium production due to screening effects in the medium was  thought of as a clean probe of the QGP and the experimental discovery of the suppression was initially celebrated as ``compelling evidence'' for the QGP. However, Elena and collaborators then showed that the same suppression could also arise due to hadronisation effects involving co-moving particles.\cite{Capella:2000zp} Also, more RHIC data showed  less suppression than theoretically predicted. To explain this also regeneration, either in the medium or at the phase boundary, has to be considered. She concluded that while the quarkonium suppression is understood qualitatively, the interplay of different competing mechanisms makes precise predictions difficult. To distinguish different models, measurements of excited states such as $\psi(2S)$ can be crucial.

Carlos Salgado presented progress in the theoretical description of jet quenching, in particular the description of in-medium parton branching. In his treatment \cite{CasalderreySolana:2012ef}, the original jet can be thought of as a set of vacuum-like subjets with a jet radius set by the medium scale. On smaller scales the jets evolve under the standard collinear evolution, while medium induced radiation destroys color-coherence over larger scales. Carlos stressed that while we have a qualitative understanding of jet-quenching, there remain open issues. In particular, measurements seem to suggest that the deviations of the quenching parameter from an ideal estimate seem to depend on global properties of the collision rather than, as one would expect,  the local properties of the medium.\cite{Andres:2016iys}

Another interesting puzzle was presented by Bj\"orn Schenke and was also main topic in the discussion session following the talks. This concerns the ridge, see Figure \ref{fig:ridges}, a structure in two-particle correlations which is long range in the rapidity difference $\Delta \eta$ and manifests itself as two bumps in the difference of azimuthal angles $\Delta\phi$. The structure can be analyzed as a function of the centrality by performing a Fourier analysis in $\Delta\phi$. The Fourier coefficients define flow coefficients $v_n$, the lowest few of which were precisely measured by ALICE \cite{ALICE:2011ab} a few years ago and agree nicely with theoretical predictions obtained using relativistic hydrodyamics.\cite{Gale:2012rq} Interestingly, the same kind of structure also arises in high-multiplicity proton-proton and proton-lead collisions, which do not involve a QGP and where a hydrodynamical description is presumably not appropriate. A possible explanation is that it is caused by inital-state correlations. The experimental results and different computations of the effects are compared and reviewed in the papers \cite{Lappi:2015vta,Dusling:2015gta}.

\begin{figure}
\centering
\includegraphics[width=0.9\linewidth]{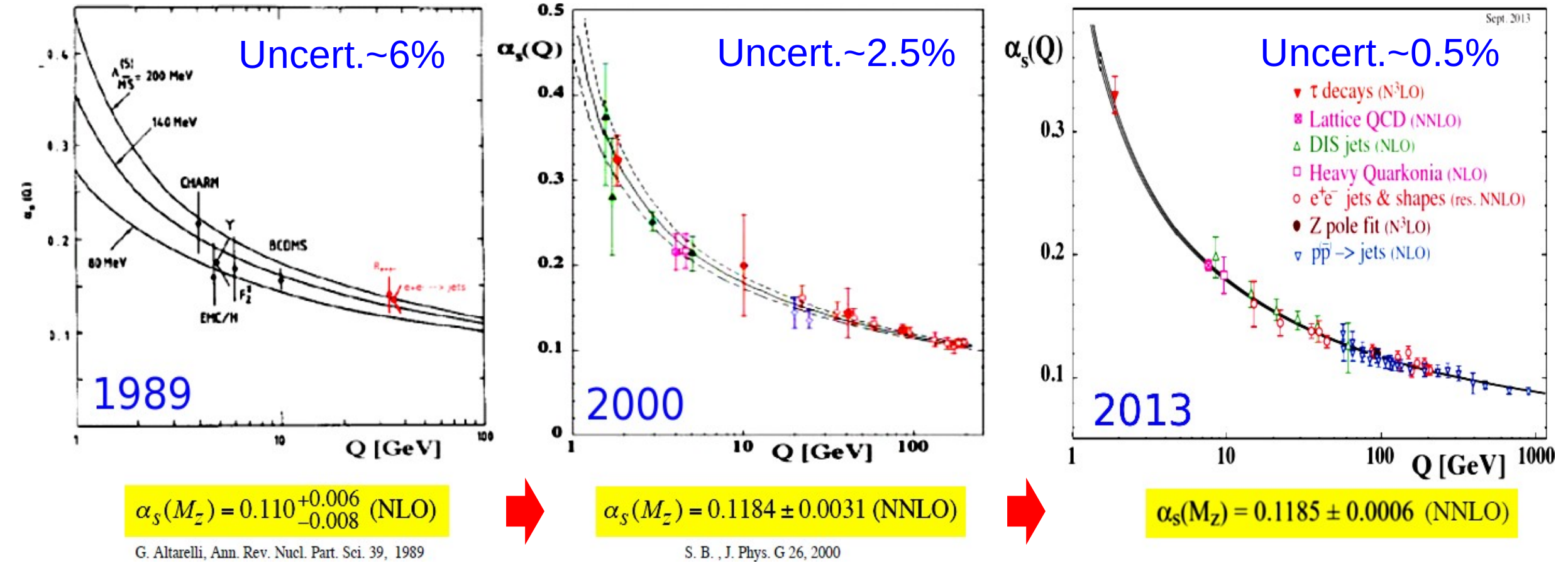}
\caption[]{Developement of the precision of $\alpha_s$ over time.}
\label{fig:alphas}
\end{figure}

\section{Conclusions}

As stressed in the introduction, sometimes it helps to look at developments over a longer period to get a perspective on how things are evolving. I enjoyed David Enterria's review \cite{d'Enterria:2016zpn} of $\alpha_s$, in which he not only discussed the current status but also showed how the precision of $\alpha_s$ increased over the years, see Figure \ref{fig:alphas}. The increase in precision is impressive and testament to our ever-increasing ability to control and compute QCD effects. The many NNLO results shown at the conference demonstrate that even in the difficult hadron-collider environment, we now have predictions at few per cent accuracy for many observables and have entered an area of precision QCD.

At the same time, QCD remains very challenging and interesting and with every increase in precision, one encounters new open questions and problems.  In the case of $\alpha_s$ the overall picture is very consistent with the QCD running, but a closer look at the different determinations reveals several tensions and difficulties. For example, the two most precise determinations (from lattice QCD \cite{McNeile:2010ji} and $e^+e^-$ event shapes variables \cite{Abbate:2010xh}) differ by about $4\sigma$. These issues are reflected in an increased uncertainty in the current PDG world average which is now $\alpha_s(M_Z) =0.1181\pm 0.0013$ corresponding to an uncertainty of  1.1\%, about twice as large as it was in the previous edition. 

Hopefully the new larger datasets at the energy frontier and in the flavor sector will eventually reveal New Physics. To explore it, our ability to control QCD effects will play a crucial role. It will be interesting to see where  the next 50 years of Moriond conferences will take us!

\newpage

\section*{Acknowledgments}

I would like to thank the organizers for the great conference and the invitation to give the summary talk and the participants for many interesting discussions. I thank Ding Yu Shao and Javier Virto for comments on the manuscript and Tom Blum for providing me with a suitable photo for the title page. A special thank you to Greg Landsberg for reminding me every day how good the weather was and how much skiing I missed while putting together summary slides. My research is supported by the Swiss National Science Foundation (SNF) under contracts 200020\_165786 and  CRSII2\_160814.

\end{document}